\documentclass[prd,aps,showpacs,amsmath,amssymb,twocolumn]{revtex4}
\usepackage{graphicx}
\usepackage{dcolumn}
\usepackage{bm}

\begin{document}

%

\let\a=\alpha      \let\b=\beta       \let\c=\chi        \let\d=\delta
\let\e=\varepsilon \let\f=\varphi     \let\g=\gamma      \let\h=\eta
\let\k=\kappa      \let\l=\lambda     \let\m=\mu
\let\o=\omega      \let\r=\varrho     \let\s=\sigma
\let\t=\tau        \let\th=\vartheta  \let\y=\upsilon    \let\x=\xi
\let\z=\zeta       \let\io=\iota      \let\vp=\varpi     \let\ro=\rho
\let\ph=\phi       \let\ep=\epsilon   \let\te=\theta
\let\n=\nu
\let\D=\Delta   \let\F=\Phi    \let\G=\Gamma  \let\L=\Lambda
\let\O=\Omega   \let\P=\Pi     \let\Ps=\Psi   \let\Si=\Sigma
\let\Th=\Theta  \let\X=\Xi     \let\Y=\Upsilon

%

%

\def\cA{{\cal A}}                \def\cB{{\cal B}}
\def\cC{{\cal C}}                \def\cD{{\cal D}}
\def\cE{{\cal E}}                \def\cF{{\cal F}}
\def\cG{{\cal G}}                \def\cH{{\cal H}}
\def\cI{{\cal I}}                \def\cJ{{\cal J}}
\def\cK{{\cal K}}                \def\cL{{\cal L}}
\def\cM{{\cal M}}                \def\cN{{\cal N}}
\def\cO{{\cal O}}                \def\cP{{\cal P}}
\def\cQ{{\cal Q}}                \def\cR{{\cal R}}
\def\cS{{\cal S}}                \def\cT{{\cal T}}
\def\cU{{\cal U}}                \def\cV{{\cal V}}
\def\cW{{\cal W}}                \def\cX{{\cal X}}
\def\cY{{\cal Y}}                \def\cZ{{\cal Z}}

\def\dbd{{$0\nu 2\beta\,$}}
%

\newcommand{\Ns}{N\hspace{-4.7mm}\not\hspace{2.7mm}}
\newcommand{\qs}{q\hspace{-3.7mm}\not\hspace{3.4mm}}
\newcommand{\ps}{p\hspace{-3.3mm}\not\hspace{1.2mm}}
\newcommand{\ks}{k\hspace{-3.3mm}\not\hspace{1.2mm}}
\newcommand{\des}{\partial\hspace{-4.mm}\not\hspace{2.5mm}}
\newcommand{\desco}{D\hspace{-4mm}\not\hspace{2mm}}



\title{\boldmath Charged Higgs contribution to \dbd decay}

\author{Namit Mahajan
}
\email{nmahajan@prl.res.in}
\affiliation{
 Theoretical Physics Division, Physical Research Laboratory, Navrangpura, Ahmedabad
380 009, India
}


\begin{abstract}
The singly charged Higgs boson contribution to \dbd is neglected on grounds of couplings involving small
masses and small nuclear matrix elements. We reconsider such contributions, but now in the light of QCD
corrections and renormalization group evolution. It is found that the charged Higgs contribution
is generically as large as (and at times significantly larger than)
the other contributions and there can be large cancellations between
contributions coming from different sources. This observation will have an important
impact on the phenomenology.
\end{abstract}

\pacs{
}
\maketitle


Charge neutrality of the neutrinos opens up the possibility of them being Majorana particles
\cite{Majorana:1937vz}. Neutrinoless 
double beta (\dbd) decay, $(A,Z)\rightarrow (A,Z+2) + 2e^-$ is an unambiguous signature
of the Majorana nature of the neutrinos. Such a process violates lepton number by two units \cite{Furry:1939qr}.
Experimental confirmation of the mixing of different neutrinos and
the fact that neutrinos are massive particles (see \cite{Tortola:2012te} for best 
fit values of the parameters) already implies physics beyond the standard model (SM).
 \dbd decay is a powerful probe of physics beyond SM since it has the potential to
 discriminate between the two hierarchies of the neutrino masses. 
 This becomes particularly important and effective in 
 the context of models which involve TeV scale particles, 
 like low scale seesaw models or low energy supersymmetric models including models with R-parity violation
  or leptoquark models. 
 More interestingly, \dbd diagrams in such low scale models can have distinctive signatures at the large hadron collider (LHC). 
  For an incomplete list discussing various aspects of \dbd decay and impact on other phenomenological
  issues see e.g. \cite{Keung:1983uu}.
  
  Experimentally, 
studies have been carried out on several nuclei (\cite{KlapdorKleingrothaus:2006ff}-
\cite{DeliaTosionbehalfoftheEXO:2014zza}). 
Only one of the experiments \cite{KlapdorKleingrothaus:2006ff}
(HM) has claimed observation of \dbd signal in $^{76}{\mathrm Ge}$. 
The half-life at $68\%$ confidence level is: $T^{0\nu}_{1/2}(^{76}{\mathrm Ge}) 
= 2.23^{+0.44}_{-0.31}\times 10^{25}\,
{\mathrm yr}$. A combination of the Kamland-Zen and EXO-200 results, both using $^{136}{\mathrm Xe}$, 
yields a lower limit on the half-life $T^{0\nu}_{1/2}(^{136}{\mathrm Xe}) > 3.4 \times 10^{25}\, {\mathrm yr}$ 
which is at variance with the HM claim. Very recently GERDA experiment reported the lower limit 
on the half-life based 
on the first phase of the experiment: $T^{0\nu}_{1/2}(^{76}{\mathrm Ge}) > 2.1 \times 10^{25}\, {\mathrm yr}$.
A combination
of all the previous limits results in a lower limit $T^{0\nu}_{1/2}(^{76}{\mathrm Ge}) > 3.0 \times 10^{25}\,
{\mathrm yr}$
at $90\%$ confidence level. The new GERDA result and the combination both are again at odds with the 
positive claim of HM.
Higher statistics in future will shed more light. One can think of comparing \dbd predictions for different
nuclei in order to study the sensitivity of of theoretical calculations on the nuclear matrix elements (NMEs) used.

 It is practically useful to separate the \dbd decay amplitude into the so called long-range 
 and short-range parts 
 (for a review of theoretical and experimental issues and the sources of uncertainties and errors, 
 see \cite{Doi:1985dx} and references therein). The long range contribution is the one arising
 when a light neutrino is exchanged while the short range part gets its name from the fact that
 the intermediate particles are all very massive and therefore the effective interaction becomes
 point-like once the heavy degrees of freedom are integrated out. 
 This distinction between the long range and the short range contributions to \dbd
 amplitude is also natural and appropriate from the point of view of renormalization and 
 evolution under renormalization group equations (RGEs). The last piece of input is the 
 non-perturbative NMEs, which are nothing but
 properly normalized matrix elements of the quark level operators sandwiched between the nucleon states.
At present, the biggest source of uncertainty stems from the NMEs, and the predictions can vary up to a 
factor of two or more depending upon the specific NMEs employed (see \cite{Simkovic:2007vu}).

Recently, for the very first time, it has been shown \cite{Mahajan:2013ixa} 
that perturbative QCD corrections to the short range part can have
an important effect on the \dbd rate. The main effect is related to the fact that QCD corrections
generate operators with colour mis-matched structure. These operators have effective couplings, called
the Wilson coefficients encoding the relevant information about the heavy degrees of freedom, which 
very roughly speaking are $1/N_c$ of the colour matched operators, $N_c$ being the number of colours.
Though accompanied by smaller coefficients, such operators when Fierz transformed can lead to
different Dirac structures whose nuclear matrix elements are way large compared to others usually 
considered. This observation is expected to have a huge impact on the phenomenological studies in a given
model. The operators generated due to the mediation of a scalar fall under this category. Another important
outcome of the QCD corrections and RG evolution to the low scale is that there is a large cancellation
between some of the colour matched and colour mis-matched operators. Operators of the form $V-A\otimes V-A$ or 
$V+A\otimes V+A$ exhibit this feature. But these are the operators that appear naturally in most
of the theories of interest, thereby making the impact of QCD corrections an important feature
that should be included in the calculation of the \dbd rate. An issue of concern is the possibility of
large cancellations among the various short range contributions, thereby significantly altering the 
limits on the masses and couplings in a given underlying model. Such cancellations (or large enhancements) will
also change the phenomenological aspects while studying the same models (applicable to low scale
models) at LHC. What is important here is the fact that such cancellations or enhancements 
do not depend on specific NMEs chosen.

A large class of models have an extended Higgs sector, popular examples being two Higgs doublet models,
supersymmetry, left-right symmetric models. In these models, apart from other particles, one has
at least one physical singly charged scalar (denoted by $H^+$) that mediates charged-current interactions.
We shall assume for the present that in each of the models considered, there are heavy right handed
neutrinos ($N$) present.
The contribution of $H^+$ to \dbd amplitude can be obtained by replacing the 
$W$'s by the $H^{\pm}$ and appropriately changing the couplings, which typically depend on the masses
of the fermions at the relevant vertex. This feature holds in all the models mentioned above.
These contributions to \dbd are simply ignored since the vertices are dependent on masses of light quarks
and/or suppression due to charged Higgs mass in the propagators. Further, the NMEs relevant for a 
contribution arising due to $H^+$ are smaller than the ones for $W$'s. All of these have prompted
one to totally discard the $H^+$ contributions. However, as argued above,
QCD corrections can change the picture completely. 
In the present note we consider the minimal left-right symmetric model (see for 
example \cite{Pati:1974yy} for the details and features of the model) for concreteness but we 
emphasize again that the features studied here remain true in all the models mentioned above.

We begin by recapitulating the essentials of the left-right model. For consistency of notation,
we follow \cite{Duka:1999uc}. The smallest gauge group implementing the left-right symmetry is 
$SU(2)_L\otimes SU(2)_R\otimes U(1)_{B-L}$. The left handed and right handed fields transform as
doublets under $SU(2)_L$ and $SU(2)_R$ respectively and therefore the two types of fields are
treated at the same footing. The model naturally contains right handed neutrinos, appearing as
a component of the doublet along with the right handed leptons. The fermionic content (and charge
assignment, with $Y=B-L$) of the model
is thus:
\begin{eqnarray}
 L_{iL} = \left(\begin{array}{c}
           \nu'_i\\ 
            \ell'_i          
          \end{array}\right)_L (2,1,-1) &,&  
     L_{iR} =  \left(\begin{array}{c}
           \nu'_i\\ 
            \ell'_i          
          \end{array}\right)_R (1,2,-1)  \\
     Q_{iL} = \left(\begin{array}{c}
           u'_i\\ 
            d'_i          
          \end{array}\right)_L (2,1,1/3) &,&  
     Q_{iR} =  \left(\begin{array}{c}
           u'_i\\ 
            d'_i          
          \end{array}\right)_R (1,2,1/3)  \nonumber    
\end{eqnarray}
The gauge couplings and the gauge fields are denoted as $g_L$, $g_R$ ($g_L=g_R=g$), $g'$, $W_L$, $W_R$, $B$.
The scalar sector of the model contains a bi-doublet and two triplets:
\begin{eqnarray}
 \phi &=& \left(\begin{array}{cc}
               \phi_1^0 & \phi_1^+\\
               \phi_2^-& \phi_2^0
              \end{array}\right) (2,2,0) \\
              \Delta_{L,R} &=& \left(\begin{array}{cc}
                                    \frac{\delta_{L,R}^+}{\sqrt{2}} & \delta_{L,R}^{++}\\
                \delta_{L,R}^0 & -\frac{\delta_{L,R}^+}{\sqrt{2}}                  
                                   \end{array}\right) \, [\Delta_{L,R} \sim (3(1),1(3),2)]\nonumber
\end{eqnarray}
The neutral components of the Higgs fields acquire vacuum expectation values (VEVs), assumed to be all real here:
\begin{equation}
 \langle\phi_{1,2}^0\rangle = \frac{\kappa_{1,2}}{\sqrt{2}},\,\, 
 \hskip 0.5cm \langle\delta_{L,R}^0\rangle = \frac{v_{L,R}}{\sqrt{2}}
\end{equation}
For what is relevant below, there are two charged gauge bosons $W_{1,2}$ with masses
\begin{equation}
 M^2_{W_{1,2}} = \frac{g^2}{4}\left(\kappa_+ + v_R^2 \mp \sqrt{v_R^4+4\kappa_1^2\kappa_2^2}\right)
\end{equation}
where $\kappa_{\pm} = \sqrt{\kappa_1^2\pm\kappa_2^2}$ and the angle $\xi$ parametrizes the mixing
between the left and right $W$ fields: $\tan 2\xi = -2\kappa_1\kappa_2/v_R^2$. In what follows,
we shall always assume $v_R >> \kappa_+$. In this limit, various expressions simplify a lot. Further,
again for simplicity we assume that $\xi$ is small and to bring out the main points relevant for the 
present study, we shall set it to zero while writing the relevant interactions. 
In the scalar sector there are $14$ physical Higgs bosons: four neutral scalars, two neutral pseudo-scalars,
two singly charged scalars ($H_{1,2}^{\pm}$) and two doubly charged scalars.  Among the singly charged scalars, one of them, $H_1^+$,
 is lighter and does not 
couple to quarks. It therefore does not participate in \dbd process. The other
singly charged scalar, $H_2^+$, is somewhat heavier but has the desired interactions. Let us assume that 
we have a TeV scale left-right model, implying that the heavy particles, including the heavy
right handed neutrinos, are all at TeV range (the heavier
among them like $H_2^+$ would be at a few TeV scale). The exact values of the masses of the particles
will depend on the details of the parameters of the model. We choose to stay somewhat generic at this
point. This has the advantage that the analysis below can be easily carried over to other
models of interest where the interactions have the same form.
For complete details about the particle spectrum, masses and interactions,
the reader is referred to \cite{Duka:1999uc}. One has the following relevant interactions:
\begin{eqnarray}
{\mathcal{L}}_{ffW} &=& \frac{g}{\sqrt{2}}\left[U_L^{CKM}(\bar{U}D)_{V-A} + 
K_L(\bar{N}\ell)_{V-A} \right]W_{1\mu}^+ \nonumber \\
&+& \frac{g}{\sqrt{2}}\left[U_R^{CKM}(\bar{U}D)_{V+A} + 
K_R(\bar{N}\ell)_{V+A} \right]W_{2\mu}^+
\end{eqnarray}
where the fields are now written in the mass basis and $U_L^{CKM}$, $U_R^{CKM}$, $K_L$ and
$K_R$ are the various mixing matrices. The fermion-charged Higgs interactions are:
\begin{eqnarray}
{\mathcal{L}}_{ffH_2^+} &=& -\bar{U}[P_L(m_uU_L^{CKM}B^+-m_dU_R^{CKM}A^+) \nonumber \\
&&+ P_R(m_uU_R^{CKM}A^+-m_dU_L^{CKM}B^+)]D \\
&+& \bar{N}_a[P_L(m_{N_a}(K_L)_{a\ell}B^+ - m_{\ell}(K_R)_{a\ell}A^+) \nonumber \\
&& + P_R((\Omega_L)_{ab}m_{N_b}(K_R)_{b\ell}A^+ - (K_L)_{a\ell}m_{\ell}B^+)]\ell_l \nonumber
\end{eqnarray}
where summation over the indices is implicit and so is the hermitian conjugate part, and
\begin{equation}
 A^+ \sim \frac{\sqrt{2}\kappa_+}{\kappa_-^2}H_2^+, \hskip 0.5cm 
 B^+ \sim \frac{2\sqrt{2}\kappa_1\kappa_2}{\kappa_+\kappa_-^2}H_2^+
\end{equation}
Note that $\kappa_-\to 0$ would give rise to singular behaviour of observables and thus 
this strict limit needs to be avoided. On the other hand, for choice of parameters, there can be 
enhancement due smaller values of $\kappa_-$. Compare the above interaction with that in
2HDMII or supersymmetry. There the charged Higgs couples to the up and down type members
of the doublets as (as an example the quarks, but the same structure will follow for the leptons
with appropriate changes):
\begin{equation}
 -\frac{\sqrt{2}}{v}V_{UD}\, \bar{U}(m_u\cot\beta P_L + m_d\tan\beta P_R)D\, H^+ + H.C.
\end{equation}
The above form is simpler than the explicit one given above for the left-right model. To gain 
a clear and quick understanding of the situation, let us momentarily work with this form. Further,
recalling that $m_u \sim m_d/2$, let us choose to take both of them to be equal for simplicity,
and denote it by $m_q$. Further, in the minimal left-right model, $U_L^{CKM}=U_R^{CKM} = V^{CKM}$.
The quark part of the \dbd amplitude will have the following structures: $(S\pm P)\otimes (S\pm P)$
and $(S\pm P)\otimes (S\mp P)$. The structures will be weighted by (in the left-right model, for
some choices of the parameters, there can be a large enhancement as discussed above - this is
not explicitly displayed for the time being though for detailed numerical analysis this will play
a crucial role)
\begin{equation}
C_H = V_{ud}^2 T_{ea}^{*2}(m_q m_{N_a})^2/(m_{N_a} m_{H_2^+}^4)
\end{equation}
where the factors $(m_q m_{N_a})^2$ and $m_{N_a} m_{H_2^+}^2$ in numerator and denominator 
respectively arise from the vertices and propagators and we have denoted the electron-heavy neutrino mixing
by $T_{ea}$. It is to be noted that the smallness of the quark mass at the vertex is compensated by the
large heavy neutrino mass. QCD corrections will lead to quark operators with colour mis-matched structure.
The weight (magnitude) of these operators after the RG evolution to the relevant low scale
is typically $0.1$-$0.5$ of the colour matched operator.

The following scalar-pseudoscalar operators are of interest for the present study:
\begin{eqnarray}
O^{SP\pm\pm}_1 &=& \bar{u_i}(1\pm\gamma_5)d_i\, \bar{u_j}(1\pm\gamma_5)d_j\, \bar{e}(1+\gamma_5)e^c \nonumber \\
O^{SP\pm\pm}_2 &=& \bar{u_i}(1\pm\gamma_5)d_j\, \bar{u_j}(1\pm\gamma_5)d_i\, \bar{e}(1+\gamma_5)e^c
\end{eqnarray}
In addition, the following tensor operators are required for RG purposes \cite{Buras:2000if}:
\begin{eqnarray}
O^{T\pm\pm}_1 &=& \bar{u_i}\sigma_{\alpha\beta}(1\pm\gamma_5)d_i\, \bar{u_j}\sigma_{\alpha\beta}
(1\pm\gamma_5)d_j\, \bar{e}(1+\gamma_5)e^c \nonumber \\
O^{T\pm\pm}_2 &=& \bar{u_i}\sigma_{\alpha\beta}(1\pm\gamma_5)d_j\, \bar{u_j}\sigma_{\alpha\beta}
(1\pm\gamma_5)d_i\, \bar{e}(1+\gamma_5)e^c
\end{eqnarray}
It is sufficient to only consider operators with $LL$ structure since the $RR$ operators will have
the same properties under QCD renormalization. Following \cite{Buras:2000if}, and adapting to the present case,
the Wilson coefficients at the low scale approximately read (in units of $C_H$, the only non-zero high scale coefficient):
\begin{eqnarray}
 C_1^{SP--} \sim 3 &,& C_2^{SP--} \sim 0.16 \\
 C_1^{T--} \sim 0.06 &,& C_2^{T--} \sim -0.17 \nonumber
\end{eqnarray}
In obtaining the above approximate numerical values of the Wilson coefficients, we have assumed that all the
heavy particles are around TeV and have used one step integrating out of heavy degrees of freedom.
Some changes are expected once the threshold effects are incorporated.
The points to be noted from above are that the coefficients $C_1^{SP\pm\pm}$ get enhanced by a factor of $3$
and the corresponding colour mis-matched coefficient is $0.16 C_H$ while the colour matched
tensor operator comes with a strength $0.06 C_H$. The operators $O^{SP\pm\pm}_2$ are now Fierz transformed
which brings the tensor structure in the picture. The sum total of this all is that there are operators 
$O^{T\pm\pm}_1$ with coefficients which are $\sim 10$-$15\%$ of $C_H$. Since the NMEs associated with the
tensor operators are way bigger than the scalar ones, one can not naively throw away the scalar 
contribution in the end. This is the big difference that is brought in by the QCD corrections and
RG evolution to the low scale. For numerical purpose we shall employ $C^{T\pm\pm}_1 = 0.12 C_H$.
The tensor-pseudotensor structure yields the following NME:
\begin{equation}
\langle {\mathcal{J}}^{\mu\nu}{\mathcal{J}}_{\mu\nu}\rangle \propto -\alpha^{SR}_2 {\mathcal{M}}_{GT,N} \label{NMET}
\end{equation}
with $\alpha^{SR}_2 \sim 9.6\frac{m_A}{m_Pm_e}$ which is much larger than the NME for scalar-pseudoscalar
operator (note the large multiplicative factor of $9.6$ appearing in $\alpha^{SR}_2 $ which will play a crucial role in
eventually enhancing the contributions):
\begin{equation}
\langle {\mathcal{J}}^{(S\pm P)}{\mathcal{J}}_{(S\pm P)}\rangle \propto - \alpha^{SR}_1 {\mathcal{M}}_{F,N} \label{NMES}
\end{equation}
with $\alpha^{SR}_1 \sim 0.145\frac{m_A}{m_Pm_e}$. This, together with the fact that
${\mathcal{M}}_{GT,N} > {\mathcal{M}}_{F,N}$ justifies the neglect of scalar contributions
to \dbd in the absence of QCD corrections.


Neglecting the contribution arising due to the doubly charged Higgs bosons, the other operators of interest are:
\begin{eqnarray}
O^{LL}_1 &=& \bar{u_i}\gamma_{\mu}(1-\gamma_5)d_i\,\bar{u_j}\gamma^{\mu}(1-\gamma_5)d_j\,\bar{e}(1+\gamma_5)e^c \nonumber \\
O^{LL}_2 &=& \bar{u_i}\gamma_{\mu}(1-\gamma_5)d_j\,\bar{u_j}\gamma^{\mu}(1-\gamma_5)d_i\,\bar{e}(1+\gamma_5)e^c \nonumber \\
O^{RR}_1 &=& \bar{u_i}\gamma_{\mu}(1+\gamma_5)d_i\,\bar{u_j}\gamma^{\mu}(1+\gamma_5)d_j\,\bar{e}(1+\gamma_5)e^c \nonumber \\
O^{RR}_2 &=& \bar{u_i}\gamma_{\mu}(1+\gamma_5)d_j\,\bar{u_j}\gamma^{\mu}(1+\gamma_5)d_i\,\bar{e}(1+\gamma_5)e^c \nonumber \\
O^{LR}_1 &=& \bar{u_i}\gamma_{\mu}(1-\gamma_5)d_i\,\bar{u_j}\gamma^{\mu}(1+\gamma_5)d_j\,\bar{e}(1+\gamma_5)e^c \nonumber \\
O^{LR}_2 &=& \bar{u_i}\gamma_{\mu}(1-\gamma_5)d_j\,\bar{u_j}\gamma^{\mu}(1+\gamma_5)d_i\,\bar{e}(1+\gamma_5)e^c
\end{eqnarray}
with the Wilson coefficients evaluated at $\mu \sim {\mathcal{O}}$(GeV) in units of the corresponding
coefficients at high scale \cite{Mahajan:2013ixa}:
\begin{eqnarray}
C^{LL,RR}_1 \sim 1.3 &,& C^{LL,RR}_2 \sim -0.6 \nonumber\\
C^{LR,RL}_1 \sim 1.1 &,& C^{LR,RL}_2 \sim 0.7
\end{eqnarray}
As noted in \cite{Mahajan:2013ixa}, there is substantial cancellation after Fierz rearrangement in the
above set of operators: $LL,\,RR$ operators effectively 
yield $C^{LL,RR}_1+C^{LL,RR}_2$ as the couplings with the same NMEs involved. 
Explicitly:
\begin{equation}
\langle {\mathcal{J}}^{(V\pm A)}{\mathcal{J}}_{(V\pm A)}\rangle \propto \frac{m_A}{m_Pm_e} 
({\mathcal{M}}_{GT,N}\,  \mp \alpha^{SR}_3 {\mathcal{M}}_{F,N}) \label{NMEV}
\end{equation}
where $\vert{\mathcal{M}}_{GT,N}\vert \sim (2-4)\vert{\mathcal{M}}_{F,N}\vert$ with
$\alpha^{SR}_3 \sim 0.63$. Thus, it is reasonable to say that the above matrix element is 
essentially governed by ${\mathcal{M}}_{GT,N}$. Let us further choose to neglect
${\mathcal{M}}_{F,N}$ which simplifies the discussion without having any appreciable impact
on phenomenology as long as the masses of the particles are all in the TeV range. If however,
the charged Higgs boson is much lighter than some of the other particles in the spectrum,
considerable care needs to be taken since the scalar operator gets enhanced at the low scale by
a large factor.

Next let us consider the situation in the minimal left-right model where all the above
operators are present. For the constraints on the model parameters see \cite{Bambhaniya:2013wza} and
references therein. In many of the analysis, it is quite common to assume $\kappa_1>>\kappa_2$. 
Instead, there is a large parameter space where $\kappa_{1,2}$ may not be this hierarchical. In such
a case, many of the constraints change. In particular, even if $m_{H_2^+}$ is $\sim 10$ TeV or so,
there is a reasonable contribution to various observables due to $\kappa_-$ appearing in
the coupling. In such a case, the contributions from the $V\pm A\otimes V\pm A$ operators and
the Fierz transformed scalar operator (yielding a tensor-pseudotensor operator) can be comparable.
Moreover, the relative signs between the two contributions can lead to large cancellations.
In that case, the short distance contribution will be dominated by the color matched
scalar operator contribution which is naively thrown away. Depending on the
couplings, particularly if there are additional sources of CP violation, there could be significant
enhancements. Either way, the phenomenological impact, i.e. effect of these on constraints on the couplings and masses,
is going to be large. 

Let us now briefly consider supersymmetric models (we assume as before that there are 
right handed neutrinos in the model), with (see for example \cite{Hirsch:1997dm}) and 
without R-parity (see for example \cite{Mohapatra:1986su}). With R-parity conserved,
the charged scalar contribution can be the largest since the charged Higgs mass is no longer
forced to be $10$ TeV or so but few hundreds of GeV. The NME for the colour mis-matched
$C_2^{SP--}$ operator will compete with other contributions, and for a charged Higgs mass
$m_{H^+} \sim 500$ GeV or so will provide the largest contribution. The situation is
more interesting in theories with R-parity violation since in such theories,
after Fierz arrangement, even without the QCD corrections, there are tensor operators. In such
a case, the low scale Wilson coefficients read:
\begin{eqnarray}
 C_1^{SP--} &\sim& 3 C_S + 0.75 C_{\lambda'},\,\,
 C_2^{SP--} \sim 0.16 C_S -2.6 C_{\lambda'} \\ 
 C_1^{T--} &\sim& 0.06 C_S + 0.74 C_{\lambda'}, \,\,
 C_2^{T--} \sim -0.17 C_S + 0.1 C_{\lambda'} \nonumber
\end{eqnarray}
where $C_S$ and $C_{\lambda'}$ denote the effective high scale coefficients of the scalar and tensor
operators at the tree level (see \cite{Mohapatra:1986su} for analytic expressions of these). 
It is very likely that the charged Higgs contribution again dominates once couplings and masses of the particles
involved satisfying all the experimental constraints are considered.

In this note, we have shown that the charged Higgs contribution to \dbd amplitude which is
usually neglected can not be ignored once QCD corrections are taken into account. In fact,
the charged Higgs contribution can finally lead to large cancellations among the short range part
or can completely overwhelm the other contributions. At any rate, this contribution
needs to be properly accounted for in detailed numerical analysis in any model beyond SM.
The impact of QCD corrections in this case is rather large and will drastically change
the constraints on the model parameters. This will also change the interplay between the limits and
constraints obtained from \dbd and model studies at LHC and/or other observables. In view of this,
it is imperative to revisit the \dbd predictions in various model in the light of these
corrections and obtain updated constraints, some of which will be totally new and unexpected since at least one
new parameter, the charged Higgs mass, will also now need to be considered.



%

\end{document}